\documentstyle[twoside,epsf]{article}

\catcode`\@=11
\long\def\@makefntext#1{
\protect\noindent \hbox to 3.2pt {\hskip-.9pt
$^{{\eightrm\@thefnmark}}$\hfil}#1\hfill}		%CAN BE USED

\def\@makefnmark{\hbox to 0pt{$^{\@thefnmark}$\hss}}	%ORIGINAL
	
\def\ps@myheadings{\let\@mkboth\@gobbletwo
\def\@oddhead{\hbox{}
\rightmark\hfil\eightrm\thepage}
\def\@oddfoot{}\def\@evenhead{\eightrm\thepage\hfil
\leftmark\hbox{}}\def\@evenfoot{}
\def\sectionmark##1{}\def\subsectionmark##1{}}

\oddsidemargin=\evensidemargin
\newcounter{sectionc}\newcounter{subsectionc}\newcounter{subsubsectionc}
\renewcommand{\section}[1] {\vspace{12pt}\addtocounter{sectionc}{1}
\setcounter{subsectionc}{0}\setcounter{subsubsectionc}{0}\noindent
	{\tenbf\thesectionc. #1}\par\vspace{5pt}}
\renewcommand{\subsection}[1] {\vspace{12pt}\addtocounter{subsectionc}{1}
\setcounter{subsubsectionc}{0}\noindent
{\bf\thesectionc.\thesubsectionc. {\kern1pt \bfit #1}}\par\vspace{5pt}}
\renewcommand{\subsubsection}[1] {\vspace{12pt}\addtocounter{subsubsectionc}{1}
	\noindent{\tenrm\thesectionc.\thesubsectionc.\thesubsubsectionc.
	{\kern1pt \tenit #1}}\par\vspace{5pt}}
\newcommand{\nonumsection}[1] {\vspace{12pt}\noindent{\tenbf #1}
	\par\vspace{5pt}}

\newcounter{appendixc}
\newcounter{subappendixc}[appendixc]
\newcounter{subsubappendixc}[subappendixc]
\renewcommand{\thesubappendixc}{\Alph{appendixc}.\arabic{subappendixc}}
\renewcommand{\thesubsubappendixc}
	{\Alph{appendixc}.\arabic{subappendixc}.\arabic{subsubappendixc}}

\renewcommand{\appendix}[1] {\vspace{12pt}
        \refstepcounter{appendixc}
        \setcounter{figure}{0}
        \setcounter{table}{0}
        \setcounter{lemma}{0}
        \setcounter{theorem}{0}
        \setcounter{corollary}{0}
        \setcounter{definition}{0}
        \setcounter{equation}{0}
        \renewcommand{\thefigure}{\Alph{appendixc}.\arabic{figure}}
        \renewcommand{\thetable}{\Alph{appendixc}.\arabic{table}}
        \renewcommand{\theappendixc}{\Alph{appendixc}}
        \renewcommand{\thelemma}{\Alph{appendixc}.\arabic{lemma}}
        \renewcommand{\thetheorem}{\Alph{appendixc}.\arabic{theorem}}
        \renewcommand{\thedefinition}{\Alph{appendixc}.\arabic{definition}}
        \renewcommand{\thecorollary}{\Alph{appendixc}.\arabic{corollary}}
        \renewcommand{\theequation}{\Alph{appendixc}.\arabic{equation}}
        \noindent{\tenbf Appendix \theappendixc #1}\par\vspace{5pt}}
\newcommand{\subappendix}[1] {\vspace{12pt}
        \refstepcounter{subappendixc}
        \noindent{\bf Appendix \thesubappendixc. {\kern1pt \bfit #1}}
	\par\vspace{5pt}}
\newcommand{\subsubappendix}[1] {\vspace{12pt}
        \refstepcounter{subsubappendixc}
        \noindent{\rm Appendix \thesubsubappendixc. {\kern1pt \tenit #1}}
	\par\vspace{5pt}}

\newcommand{\textlineskip}{\baselineskip=13pt}
\newcommand{\smalllineskip}{\baselineskip=10pt}

\newcommand{\copyrightheading}[1]
	{\vspace*{-2.5cm}\smalllineskip{\flushleft
	{\footnotesize Quantum Information and Computation, Vol.~1, No.~0 (2001) 000--000 #1}\\
	{\footnotesize \copyright\kern2pt Rinton Press}\\
	 }}

\newcommand{\publisher}[2]{{\begin{center}\footnotesize\smalllineskip
	Received #1\\
	Revised #2
	\end{center}
	}}

\def\abstracts#1#2#3{{
	\centering{\begin{minipage}{4.5in}\footnotesize\baselineskip=10pt
	\parindent=0pt #1\par
	\parindent=15pt #2\par
	\parindent=15pt #3
	\end{minipage}}\par}}

\def\keywords#1{{
	\centering{\begin{minipage}{4.5in}\footnotesize\baselineskip=10pt
	{\footnotesize\it Keywords}\/: #1
	 \end{minipage}}\par}}
\def\communicate#1{{
	\centering{\begin{minipage}{4.5in}\footnotesize\baselineskip=10pt
	{\footnotesize\it Communicated by}\/: #1
	 \end{minipage}}\par}}

\renewenvironment{thebibliography}[1]
        {\frenchspacing
	 \ninerm\baselineskip=11pt
         \begin{list}{\arabic{enumi}.}
        {\usecounter{enumi}\setlength{\parsep}{0pt}
	 \setlength{\leftmargin 12.7pt}{\rightmargin 0pt}%FOR 1--9 ITEMS
         \setlength{\itemsep}{0pt} \settowidth
	{\labelwidth}{#1.}\sloppy}}{\end{list}}

\newcounter{itemlistc}
\newcounter{romanlistc}
\newcounter{alphlistc}
\newcounter{arabiclistc}

\newcommand{\fcaption}[1]{
        \refstepcounter{figure}
        \setbox\@tempboxa = \hbox{\footnotesize Fig.~\thefigure. #1}
        \ifdim \wd\@tempboxa > 5in
           {\begin{center}
        \parbox{5in}{\footnotesize\smalllineskip Fig.~\thefigure. #1}
            \end{center}}
        \else
             {\begin{center}
             {\footnotesize Fig.~\thefigure. #1}
              \end{center}}
        \fi}

\newcommand{\tcaption}[1]{
        \refstepcounter{table}
        \setbox\@tempboxa = \hbox{\footnotesize Table~\thetable. #1}
        \ifdim \wd\@tempboxa > 5in
           {\begin{center}
        \parbox{5in}{\footnotesize\smalllineskip Table~\thetable. #1}
            \end{center}}
        \else
             {\begin{center}
             {\footnotesize Table~\thetable. #1}
              \end{center}}
        \fi}

\def\pmb#1{\setbox0=\hbox{#1}
	\kern-.025em\copy0\kern-\wd0
	\kern.05em\copy0\kern-\wd0
	\kern-.025em\raise.0433em\box0}

\def\fnt#1#2{\footnotetext{\kern-.3em
	{$^{\mbox{\scriptsize #1}}$}{#2}}}

\def\fpage#1{\begingroup
\voffset=.3in
\thispagestyle{empty}\begin{table}[b]\centerline{\footnotesize #1}
	\end{table}\endgroup}

\def\runninghead#1#2{\pagestyle{myheadings}
\markboth{{\protect\footnotesize\it{\quad #1}}\hfill}
{\hfill{\protect\footnotesize\it{#2\quad}}}}
\font\tenrm=cmr10
\font\tenit=cmti10
\font\tenbf=cmbx10
\font\bfit=cmbxti10 at 10pt
\font\ninerm=cmr9
\font\nineit=cmti9
\font\ninebf=cmbx9
\font\eightrm=cmr8

\font\sevenrm=cmr7

\def\FigName{figure}%
\newbox\captionbox
\long\def\@makecaption#1#2{%
  \ifx\FigName\@captype
    \vskip\abovecaptionskip
    \setbox\tempbox\hbox{{\figurecaptionfont #1\hskip1em #2}}
	\ifdim\wd\tempbox< 28pc
	\centerline{\box\tempbox}
	\else
	{\figurecaptionfont #1\hskip1em #2\par}
\fi\else
  	\setbox\tempbox\hbox{{\tablecaptionfont #1\hskip1em #2}}
 	\ifdim\wd\tempbox< 28pc
	\centerline{\box\tempbox}
	\else
	{\tablecaptionfont #1\hskip1em #2\par}%
	\fi
 \vskip\belowcaptionskip
 \fi}
\InputIfFileExists{psfig.sty}
{\typeout{^^Jpsfig.sty inputed...ok}}{\typeout{^^JWarning: psfig.sty could be be found.^^J}}
\InputIfFileExists{epsfsafe.tex}
{\typeout{^^Jepsfsafe.tex inputed...ok}}
			{\typeout{^^JWarning: epsfsafe.tex could not be found.^^J}}
\InputIfFileExists{epsfig.sty}
{\typeout{^^Jepsfig.sty inputed...ok}}{\typeout{^^JWarning: epsfig.sty could not be found.^^J}}
\InputIfFileExists{epsf.sty}
{\typeout{^^Jepsf.sty inputed...ok}}{\typeout{^^JWarning: epsf.sty could not be found.^^J}}%
\def\fps@figure{tbp}
\def\ftype@figure{1}
\def\ext@figure{lof}
\def\fnum@figure{Fig.\ \thefigure}
\def\qed{\hbox{${\vcenter{\vbox{	          %HOLLOW SQUARE
   \hrule height 0.4pt\hbox{\vrule width 0.4pt height 6pt
   \kern5pt\vrule width 0.4pt}\hrule height 0.4pt}}}$}}

\begin{document}
\setlength{\textheight}{8.0truein}    %FOR 2ND PAGE ONWARDS

\runninghead{Separability and Entanglement of Spin $1$ Particle}
            {V.I. Man'ko, L.A. Markovich}

\normalsize\textlineskip
\thispagestyle{empty}
\setcounter{page}{1}

\copyrightheading{}	%	{Vol.~1, No.~0 (2001) 000--000}

\vspace*{0.88truein}

\fpage{1}
\centerline{\bf
%%%%%%%%%%%%%%%%%%%%%
%Put in titiles here
%%%%%%%%%%%%%%%%%%%%%
SEPARABILITY AND ENTANGLEMENT OF SPIN $1$ PARTICLE}
\vspace*{0.37truein}
\centerline{\footnotesize
V.I. MAN'KO\footnote{manko@sci.lebedev.ru}}
\vspace*{0.015truein}
%\centerline{\footnotesize\it University Department, University Name, Address}
\baselineskip=10pt
\centerline{\footnotesize\it P.N. Lebedev Physical Institute, Russian Academy of Sciences,
Leninskii Prospect 53, Moscow 119991, Russia}
\vspace*{10pt}
\centerline{\footnotesize
L.A. MARKOVICH\footnote{kimo1@mail.ru}}
\vspace*{0.015truein}
%\centerline{\footnotesize\it Group, Laboratory, Address}
\baselineskip=10pt
\centerline{\footnotesize\it Moscow Institute of Physics and Technology, State University, Moscow, Russia}
\vspace*{0.225truein}
\publisher{(received date)}{(revised date)}

\vspace*{0.21truein}
\abstracts{
%%%%%%%%%%%%%%%%%%%%
% put abstract here
%%%%%%%%%%%%%%%%%%%%
We define the separability and entanglement notion for particle with spin $s=1$. We consider two cases. In the first the particle is composed of two fermions with $s_1=1/2$ and $s_2=1/2$. In the second case the state is the qutrit state which is not composed system. The notion of negativity and concurrence is defined for the qutrit state. The concurrence and negativity of entangled and separable qutrit states determined by the parameters of the density matrix are explicitly calculated. The maximum entanglement of the qutrit state is observed for maximum values of non diagonal matrix elements of the density matrix. New entropic inequalities for the density matrix of the qutrit state are obtained.
}{}{}

\vspace*{10pt}
\vspace*{3pt}
\communicate{to be filled by the Editorial}

\vspace*{1pt}\textlineskip	%) USE THIS MEASUREMENT WHEN THERE IS
\section{Introduction}	        %) A SECTION HEADING
\vspace*{-0.5pt}
\noindent
\par The notion of entanglement \cite{schredinger:35} is considered for composite systems which have subsystems.
Two particles with spins $1/2$ provide an example of such system. The composite systems have correlations of degrees of freedom of the subsystems. Notion of single particle entanglement was studied for example in \cite{Can:2005}. 
Recently the possibility to study quantum correlations in the noncomposite systems were discussed \cite{Klyachko:2008,Manko:2014,Chernega:2014,Markovich:2014}. The presence of quantum correlations in noncomposite systems
is detected by different kinds of inequalities like Bell inequalities \cite{Clauser:1969,Lieb:1974,Wehner:2010} which are violated for entangled states \cite{Cirelson:1980} and inequalities for von Neumann entropies
and informations determined by density matrices of the system and its subsystems.
As it was pointed out in \cite{Markovich2:2014} analogues inequalities  can be obtained for systems without subsystems.
\par The aim of our work is to study the notion of entanglement properties introducing the concurrence and negativity of $3\times3$ density matrix. The entropic inequalities of the concrete system with spin $1$
(qutrit) when the system corresponds to two fermions in symmetric spin state are obtained. Analogous entropic inequalities are studied for non composite system (tree-level atom).
We get the analog of subadditivity condition for the von Neumann entropy of the system state
and for two entropies which describe properties of two "qubit" density matrices. These density matrices are  obtained by means of positive map from the density matrix of the system.
This positive map corresponds to map of indices labeling  basis vectors in the Hilbert space of the system states. For example, we use the map
$|\uparrow\uparrow>\leftrightarrow|1,1>\leftrightarrow 1$
for which the usual notation  for spin state projections $+1/2$ $+1/2$ is denoted as $|11>$. Other basis vectors are labeled by analogously. The analog of negativity and concurrence was introduced for this state.
\par The outline of the paper is as follows. In Section 2 standard entropic inequalities known for two-qubit systems are reviewed. In Section 3
the notion of separability and entanglement is described for the composite system of two spin $1/2$ particles. In Section  4 the spin $1$ system is considered as a
symmetric spin state of two spin $1/2$ particles. It is studied as one qutrit and the entanglement notion for this qutrit is discussed together with new entropic inequalities.
Prospectives and conclusion are presented in Section 5.

\vspace*{1pt}\textlineskip	%) USE THIS MEASUREMENT WHEN THERE IS
\section{Review on Two-qubit Entropic Inequalities}	\label{sec:1}        %) A SECTION HEADING
\vspace*{-0.5pt}
\noindent
The notion of entanglement is best studied on the example of a  pair of qubits, each defined on two dimensional Hilbert space. The partial Peres transposition condition gives a criterion of entanglement for this bipartite system. However, for more complicated systems the notion of entanglement is stile an open problem. In this paper we would like to study the entanglement for the qutrit state using known technics for two-qubit system.
\\To this end, let us define a quantum system on the Hilbert space. The density matrix of a system state in a four-dimensional Hilbert space $\mathcal{H}$ is
\begin{eqnarray}\rho={\left(
                                 \begin{array}{cccc}
                                   \rho_{11}& \rho_{12}& \rho_{13}& \rho_{14}\\
                                   \rho_{21}& \rho_{22}& \rho_{23}& \rho_{24}\\
                                   \rho_{31}& \rho_{32}& \rho_{33}& \rho_{34}\\
                                   \rho_{41}& \rho_{42}& \rho_{43}& \rho_{44}\\
                                 \end{array}
                               \right)}\,. \label{1}
                               \end{eqnarray}
The latter matrix has standard properties of the density matrix:  $\rho=\rho^{\dagger}$, $Tr\rho=1$  and its eigenvalues are nonnegative.
Next, we define the invertible map of indices $1\leftrightarrow 1/2~1/2$; $2\leftrightarrow1/2~-1/2$; $3\leftrightarrow-1/2~1/2$; $4\leftrightarrow-1/2~-1/2$. Applying this mapping to the density matrix $\rho$ one can write
\begin{eqnarray}\rho_{1/2}={\left(
                                 \begin{array}{cccc}
                                   \rho_{\frac{1}{2}~\frac{1}{2},\frac{1}{2}~\frac{1}{2}}& \rho_{\frac{1}{2}~\frac{1}{2},\frac{1}{2}~-\frac{1}{2}}& \rho_{\frac{1}{2}~\frac{1}{2},-\frac{1}{2}~\frac{1}{2}}& \rho_{\frac{1}{2}~\frac{1}{2},-\frac{1}{2}~-\frac{1}{2}}\\
                                   \rho_{\frac{1}{2}~-\frac{1}{2},\frac{1}{2}~\frac{1}{2}}& \rho_{\frac{1}{2}~-\frac{1}{2},\frac{1}{2}~-\frac{1}{2}}& \rho_{\frac{1}{2}~-\frac{1}{2},-\frac{1}{2}~\frac{1}{2}}& \rho_{\frac{1}{2}~-\frac{1}{2},-\frac{1}{2}~-\frac{1}{2}}\\
                                   \rho_{-\frac{1}{2}~\frac{1}{2},\frac{1}{2}~\frac{1}{2}}& \rho_{-\frac{1}{2}~\frac{1}{2},\frac{1}{2}~-\frac{1}{2}}& \rho_{-\frac{1}{2}~\frac{1}{2},-\frac{1}{2}~\frac{1}{2}}& \rho_{-\frac{1}{2}~\frac{1}{2},-\frac{1}{2}~-\frac{1}{2}}\\
                                   \rho_{-\frac{1}{2}~-\frac{1}{2},\frac{1}{2}~\frac{1}{2}}& \rho_{-\frac{1}{2}~-\frac{1}{2},\frac{1}{2}~-\frac{1}{2}}& \rho_{-\frac{1}{2}~-\frac{1}{2},-\frac{1}{2}~\frac{1}{2}}& \rho_{-\frac{1}{2}~-\frac{1}{2},-\frac{1}{2}~-\frac{1}{2}}\\
                                 \end{array}
                               \right)}\,. \label{2}
                               \end{eqnarray}
The basis $\overline{e_1}=|\uparrow\uparrow\rangle$, $\overline{e_2}=|\uparrow\downarrow\rangle$, $\overline{e_3}=|\downarrow\uparrow\rangle$,
 $\overline{e_4}=|\downarrow\downarrow\rangle$ is defined for this two-qubit state. Let us consider two subsystems on spaces $\mathcal{H}$$^{1}$ and $\mathcal{H}$$^{2}$ such that
$\mathcal{H}=\mathcal{H}$$^1\otimes\mathcal{H}$$^2$.
Reduced density operators $\rho_1$, $\rho_2$ are defined as partial traces  of $\rho_{1/2}$. Result operators are operators on  spaces $\mathcal{H}$$^1$ and $\mathcal{H}$$^2$, respectively.
Thus the reduced density matrices of the first and the second qubit are defined as
\begin{eqnarray}\rho_1={\left(
                              \begin{array}{cc}
                                \rho_{11}+\rho_{22} & \rho_{13}+\rho_{24} \\
                               \rho_{31}+\rho_{42} &  \rho_{33}+\rho_{44} \\
                              \end{array}
                            \right),\quad
                             \rho_2=\left(
                              \begin{array}{cc}
                                \rho_{11}+\rho_{33} & \rho_{12}+\rho_{34} \\
                               \rho_{21}+\rho_{43} &  \rho_{22}+\rho_{44} \\
                              \end{array}
                            \right)}\,. \label{5}
\end{eqnarray}
In terms of the density matrix for two-qubit state (\ref{2}), it can be rewritten as following
\begin{eqnarray}\rho_1={\left(
                            \begin{array}{cc}
                              \rho_{\frac{1}{2}~\frac{1}{2},\frac{1}{2}~\frac{1}{2}}+ \rho_{\frac{1}{2}~-\frac{1}{2},\frac{1}{2}~-\frac{1}{2}}& \rho_{\frac{1}{2}~\frac{1}{2},-\frac{1}{2}~\frac{1}{2}}+ \rho_{\frac{1}{2}~-\frac{1}{2},-\frac{1}{2}~-\frac{1}{2}}\\
                               \rho_{-\frac{1}{2}~\frac{1}{2},\frac{1}{2}~\frac{1}{2}}+\rho_{-\frac{1}{2}~-\frac{1}{2},\frac{1}{2}~-\frac{1}{2}}&\rho_{-\frac{1}{2}~\frac{1}{2},-\frac{1}{2}~\frac{1}{2}}+\rho_{-\frac{1}{2}~-\frac{1}{2},-\frac{1}{2}~-\frac{1}{2}} \\
                            \end{array}
                          \right)}\, , \label{7}
\end{eqnarray}
\begin{eqnarray}
                          \rho_2={\left(
                            \begin{array}{cc}
                              \rho_{\frac{1}{2}~\frac{1}{2},\frac{1}{2}~\frac{1}{2}}+\rho_{-\frac{1}{2}~\frac{1}{2},-\frac{1}{2}~\frac{1}{2}} & \rho_{\frac{1}{2}~\frac{1}{2},\frac{1}{2}~-\frac{1}{2}}+ \rho_{-\frac{1}{2}~\frac{1}{2},-\frac{1}{2}~-\frac{1}{2}}\\
                              \rho_{\frac{1}{2}~-\frac{1}{2},\frac{1}{2}~\frac{1}{2}}+\rho_{-\frac{1}{2}~-\frac{1}{2},-\frac{1}{2}~\frac{1}{2}} & \rho_{\frac{1}{2}~-\frac{1}{2},\frac{1}{2}~-\frac{1}{2}}+ \rho_{-\frac{1}{2}~-\frac{1}{2},-\frac{1}{2}~-\frac{1}{2}}\\
                            \end{array}
                          \right)}\,. \label{77}
\end{eqnarray}
Hence, the von Neumann entropies for qubit subsystems and for the hall two-qubit system can be obtained as
\begin{eqnarray}S_1 ={ -Tr\rho_1\ln\rho_1,\quad
S_2 =-Tr\rho_2\ln\rho_2,\quad
S_{12} =-Tr\rho_{1/2}\ln\rho_{1/2}}\,. \label{4}
\end{eqnarray}
The quantum information is defined as the difference of the sum of the entropies of the first and the second qubit states and the entropy of the two-qubit state, i.e.
\begin{eqnarray}I_q={S_1 +S_2-S_{12}}\,. \label{12}
\end{eqnarray}
Obviously, the quantum information satisfies the inequality $I_q\geq 0$. Hence, we get
\begin{eqnarray}{-Tr\rho_1\ln\rho_1-Tr\rho_2\ln\rho_2\geq-Tr\rho_{1/2}\ln\rho_{1/2}}\,. \label{3}
\end{eqnarray}
The inequality (\ref{3}) is called the subadditivity condition  of the system of two qubits.

\vspace*{1pt}\textlineskip	%) USE THIS MEASUREMENT WHEN THERE IS
\section{Separability and Entanglement}	\label{sec:2}        %) A SECTION HEADING
\vspace*{-0.5pt}
\noindent
Let us define the composite quantum system on the Hilbert space $\mathcal{H}^{AB\ldots}$ and its subsystems on
$\mathcal{H}^{A}$, $\mathcal{H}^{B}$ etc. It is postulated that the composite system of these subsystems is represented by their direct product $\mathcal{H}^{AB\ldots}=\mathcal{H}^{A}\otimes\mathcal{H}^{B}\otimes\ldots$.
The following definitions are introduced in \cite{Werner:1989}.
\par The composite state is called correlated if and only if its density operator $\rho^{AB\ldots}$ cannot be written as a product operator \begin{eqnarray*}\rho^{AB\ldots}&\neq&\rho_i^{A}\otimes\rho_i^{B}\otimes\ldots.\end{eqnarray*}
\par A state is called separable if and only if the density operator of the composite system $\rho^{AB}$ can be written as
\begin{eqnarray*}\rho^{AB\ldots}&=&\sum\limits_{i=1}^{n}p_i\rho_i^{A}\otimes\rho_i^{B}\otimes\ldots,\quad \sum\limits_{i=1}^{n}p_i=1.
\end{eqnarray*}
The separable state has no quantum entanglement.
\par A state is called entangled  if and only if the state is correlated and not separable
\begin{eqnarray*}\rho^{AB\ldots}&\neq&\sum\limits_{i=1}^{n}p_i\rho_i^{A}\otimes\rho_i^{B}\otimes\ldots,\quad \sum\limits_{i=1}^{n}p_i=1.
\end{eqnarray*}
Below, we will consider the two-qubit system described in Section 2. Thus, latter definitions can be rewritten in terms of matrices (\ref{2}) and (\ref{7}). For example the two-qubit state is correlated if and only if
\begin{eqnarray*}\rho_{1/2}&\neq&\rho_{1}\otimes\rho_{2}.
\end{eqnarray*}

\vspace*{1pt}\textlineskip	%) USE THIS MEASUREMENT WHEN THERE IS
\section{Results}\label{sec:3}	        %) A SECTION HEADING
\vspace*{-0.5pt}
\noindent
\subsection{Entropic inequality for qutrit}
\begin{figure} [htbp]
\vspace*{13pt}
\centerline{\psfig{file=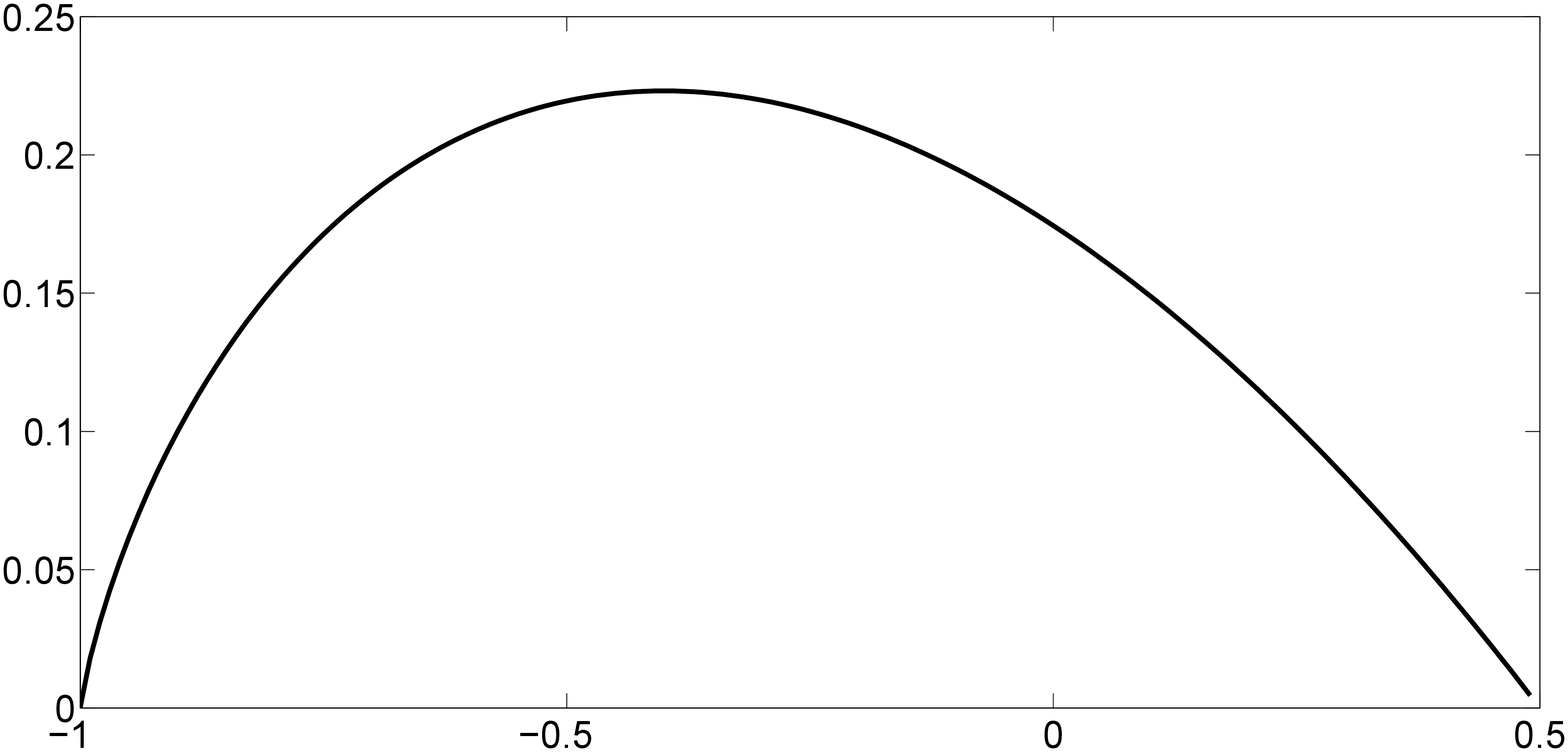, width=8.2cm}} %100 percent
\vspace*{13pt}
\fcaption{\label{fig:2}
$I_q$ (Y-axes) for the state with the density matrix (\ref{16}), $-1\leq b\leq1/2$ (X-axes).}
\end{figure}

\noindent
The system with spin $1$ (qutrit) is defined on the tree dimensional Hilbert space. Hence, we define the $3\times3$ density matrix
\begin{eqnarray}\rho={\left(
                                 \begin{array}{ccc}
                                   \rho_{11}& \rho_{12}& \rho_{13}\\
                                   \rho_{21}& \rho_{22}& \rho_{23}\\
                                   \rho_{31}& \rho_{32}& \rho_{33}\\
                                 \end{array}
                               \right),\quad \rho=\rho^{\dagger},\quad Tr\rho=1}\,. \label{17}
                               \end{eqnarray}
 with nonnegative eigenvalues. We shall use results known for the two-qubit system (see Section 2). To this end, we
add the matrix (\ref{17}) with zero row and column so that its dimension becomes $4\times4$, i.e.
\begin{eqnarray}\widetilde{\rho}={\left(
                                 \begin{array}{cccc}
                                   \rho_{11}& \rho_{12}& \rho_{13}& 0\\
                                   \rho_{21}& \rho_{22}& \rho_{23}& 0\\
                                   \rho_{31}& \rho_{32}& \rho_{33}& 0\\
                                   0& 0& 0&0\\
                                 \end{array}
                               \right)}\,. \label{11}
                               \end{eqnarray}
                               It is obvious, that latter procedure does not change the properties of the density matrix (\ref{17}), apart of the fours zero eigenvalue appears.
The latter density matrix can describe a state in a four-dimensional Hilbert space $\mathcal{H}$. Then analogously to (\ref{5}) we can write
two reduced density matrices which can be considered as density matrices of artificial "qubit" states
  \begin{eqnarray}\widetilde{\rho_1}={\left(
                              \begin{array}{cc}
                                \rho_{11}+\rho_{22} & \rho_{13} \\
                               \rho_{31} &  \rho_{33} \\
                              \end{array}
                            \right),\quad
                          \widetilde{\rho_2}=\left(
                              \begin{array}{cc}
                                \rho_{11}+\rho_{33} & \rho_{12} \\
                               \rho_{21} &  \rho_{22} \\
                              \end{array}
                            \right)}\,. \label{6}
\end{eqnarray}
Hence, like in Section 3, the von Neumann entropies for the qutrit can be obtained as
\begin{eqnarray}\label{4}\widetilde{S_1} &=& -Tr\widetilde{\rho_1}\ln\widetilde{\rho_1} = - Tr\left(
                              \begin{array}{cc}
                                \rho_{11}+\rho_{22} & \rho_{13} \\
                               \rho_{31} &  \rho_{33} \\
                              \end{array}
                            \right)\ln\left(
                              \begin{array}{cc}
                                \rho_{11}+\rho_{22} & \rho_{13} \\
                               \rho_{31} &  \rho_{33} \\
                              \end{array}
                            \right),\\\nonumber
\widetilde{S_2} &=&-Tr\widetilde{\rho_2}\ln\widetilde{\rho_2}= - Tr\left(
                              \begin{array}{cc}
                                \rho_{11}+\rho_{33} & \rho_{12} \\
                               \rho_{31} &  \rho_{22} \\
                              \end{array}
                            \right)\ln\left(
                              \begin{array}{cc}
                                \rho_{11}+\rho_{33} & \rho_{12} \\
                               \rho_{21} &  \rho_{22} \\
                              \end{array}
                            \right),\\\nonumber
\widetilde{S_{12} }&=&-Tr\widetilde{\rho}\ln\widetilde{\rho}.
\end{eqnarray}
Note that the $\widetilde{S_1}$, $\widetilde{S_2}$ are not the entropies of subsystems because  the system with spin $1$ does not contain subsystems.
Analogically to Section 2 the quantum information is defined as
\begin{eqnarray}I_q={\widetilde{S_{1}} +\widetilde{S_{2}}-\widetilde{S_{12}}}\,. \label{12}
\end{eqnarray}
The entropy functions can be expressed in terms of the eigenvalues. For example, we have
  \begin{eqnarray*}
\widetilde{S_{12} }&=&-\sum\limits_{i=1}^{4}\lambda_i\ln\lambda_i
\end{eqnarray*}
where $\lambda_i,\quad i=1,2,3,4$ are the eigenvalues of the density matrix (\ref{11}).
Obviously, the quantum information satisfies the inequality $I_q\geq 0$. Hence, using (\ref{3}), we can write for the matrix (\ref{11}) the following inequality \cite{Chernega:2013}
\begin{eqnarray}\label{23}I_q&=&- Tr\left(
                              \begin{array}{cc}
                                \rho_{11}+\rho_{22} & \rho_{13} \\
                               \rho_{31} &  \rho_{33} \\
                              \end{array}
                            \right)\ln\left(
                              \begin{array}{cc}
                                \rho_{11}+\rho_{22} & \rho_{13} \\
                               \rho_{31} &  \rho_{33} \\
                              \end{array}
                            \right)\\\nonumber
&-&Tr\left(
                              \begin{array}{cc}
                                \rho_{11}+\rho_{33} & \rho_{12} \\
                               \rho_{21} &  \rho_{22} \\
                              \end{array}
                            \right)\ln\left(
                              \begin{array}{cc}
                                \rho_{11}+\rho_{33} & \rho_{12} \\
                               \rho_{21} &  \rho_{22} \\
                              \end{array}
                            \right)+\sum\limits_{i=1}^{4}\lambda_i\ln\lambda_i\geq 0.
\end{eqnarray}
The inequality (\ref{23}) is the subadditivity condition of the system with spin $1$, which does not contain subsystems.
\\ As an example of a qutrit density matrix we can take the following one
\begin{eqnarray}\rho={\frac{1}{3}\left(
                                 \begin{array}{ccc}
                                   1+b& 0& 0\\
                                   0& 1+b& 0\\
                                   0& 0& 1-2b\\
                                 \end{array}
                               \right)}\,,\label{16}
                               \end{eqnarray}
where from the positivity condition it can be deduced that $-1\leq b\leq1/2$. Adding this matrix with zeros, like in (\ref{11}), we can rewrite entropies (\ref{4}) for a qutrit state. Hence, the inequality (\ref{23})  is determined by
\begin{eqnarray*}I_q&=&\frac{1}{3}\Bigg(\left(\ln\left(\frac{b+1}{3}\right)+\ln\left(\frac{2-b}{3}\right)+\ln\left(\frac{2b+2}{3}\right)\right)b\\
&+&
\left(\ln\left(\frac{b+1}{3}\right)+\ln\left(\frac{2-b}{3}\right)+\ln\left(\frac{2b+2}{3}\right)\right)\Bigg)\geq0.
\end{eqnarray*}
In Figure \ref{fig:2} the latter inequality is shown for different value of the parameter $b$.

\subsection{Negativity and concurrence}
\begin{figure} [htbp]
\vspace*{13pt}
\centerline{\psfig{file=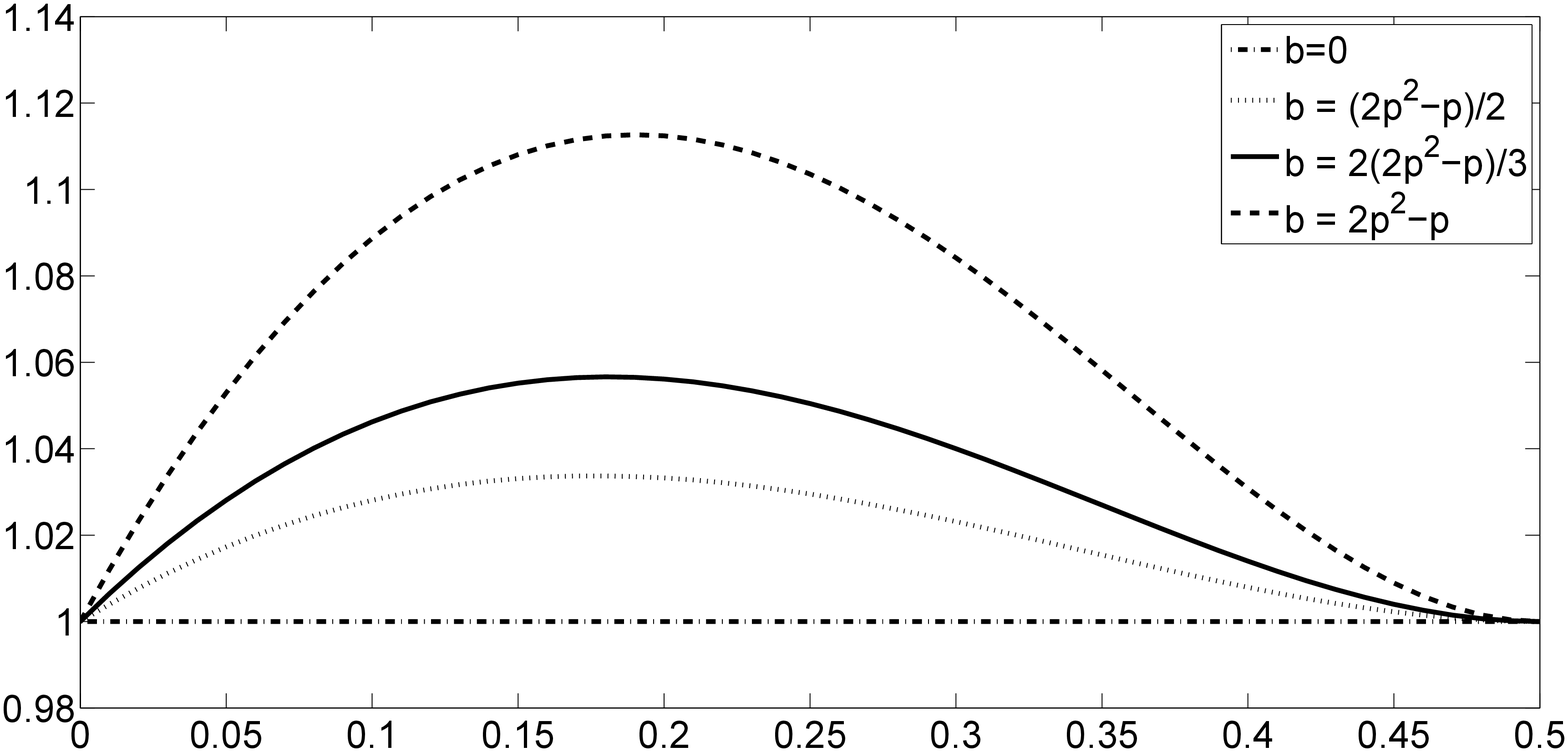, width=8.2cm}} %100 percent
\vspace*{13pt}
\fcaption{\label{fig:1}
Negativity for matrix (\ref{22}) (Y-axes), $0<p<0.5$ (X-axes).}
\end{figure}

\noindent
In \cite{Hill:1997,Wootters:1998} the notion of concurrence for two-qubit system was introduced. In this section we will define the concurrence for the noncomposite system on the example of the single qutrit.
 Let us introduce the analog of positive partial transpose operation which is known for bipartite
system for the qutrit state. We consider the density matrix (\ref{17}) zero-added as in (\ref{11}). Then the ppt matrix with eigenvalues $\lambda^{ppt}$ is
determined by
                               \begin{eqnarray}
\rho^{ppt}={\left(
                     \begin{array}{cccc}
                       \rho_{11} & \rho_{21} & \rho_{13} & \rho_{23}\\
                       \rho_{12} & \rho_{22}& 0 & 0 \\
                      \rho_{31}& 0& \rho_{33} & 0 \\
                      \rho_{32} & 0 & 0 & 0 \\
                     \end{array}
                   \right)}\,, \label{18}
\end{eqnarray}
 In the case when the following inequality
\begin{eqnarray*}&&|\lambda_1^{ppt}|+|\lambda_2^{ppt}|+|\lambda_3^{ppt}|+|\lambda_4^{ppt}|>1
\end{eqnarray*}
holds, we can interpret the sum in the left-hand side of this inequality as negativity parameter characterizing the state of  the qutrit.
As an example, we can take the following two-parametric matrix
\begin{eqnarray}\rho={\left(
                                 \begin{array}{cccc}
                                   p& 0& 0\\
                                   0& 1-2p& b\\
                                   0& b& p\\
                                 \end{array}
                               \right)}\,. \label{22}
                               \end{eqnarray}
Adding the latter matrix with zero row and column as in (\ref{11}) and doing ppt operation, we can write the negativity as
\begin{eqnarray*}&&|1-2p|+|p|+\Bigg|\frac{p}{2}-\frac{\sqrt{4b^2+p^2}}{2}\Bigg|+\Bigg|\frac{p}{2}+\frac{\sqrt{4b^2+p^2}}{2}\Bigg|>1
\end{eqnarray*}
where due to the fact that eigenvalues must be nonnegative, parameters must satisfy conditions  $0<p<0.5$, $2p^2-p<b<0$. In Figure \ref{fig:1} the negativity is shown for various parameter $b$.
\par
\begin{figure} [htbp]
\vspace*{13pt}
\centerline{\psfig{file=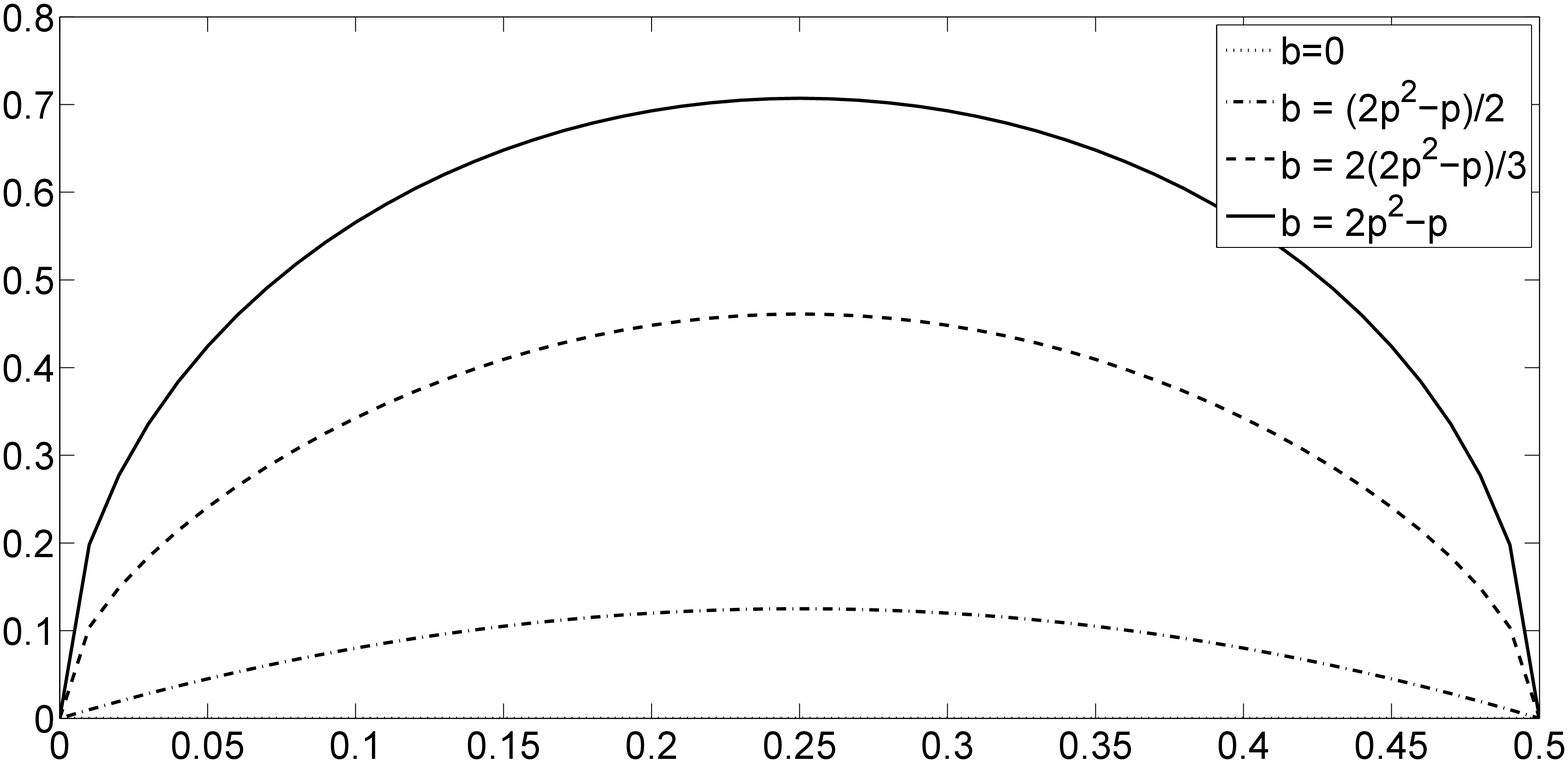, width=8.2cm}} %100 percent
\vspace*{13pt}
\fcaption{\label{fig:3}
Concurrence for matrix (\ref{22}) (Y-axes), $0<p<0.5$ (X-axes).}
\end{figure}

 The concurrence is defined by \cite{Wootters:1998}
 \begin{eqnarray}C(\rho)= {\max\{0,\sqrt{\lambda_1}-\sqrt{\lambda_2}-\sqrt{\lambda_3}-\sqrt{\lambda_4}\}}\,, \label{C}
 \end{eqnarray}
 where $\lambda_i,i=1,2,3,4$ are the eigenvalues of matrix $\widetilde{\rho}\rho^C$ in decreasing order.  We use the definition of concurrence which
 is analogue to concurrence for two-qubit system. The matrix $\rho^C$
 is obtained by spin flip operation on the qudit density matrix
                               \begin{eqnarray}
\rho^C={(\sigma_y\otimes\sigma_y)\widetilde{\rho}^{\ast}(\sigma_y\otimes\sigma_y)}\,, \label{20}
\end{eqnarray}
where $\sigma_y$ is a Pauli matrix and
\begin{eqnarray*}(\sigma_y\otimes\sigma_y)&=&\left(
                     \begin{array}{cccc}
                       0 & 0 & 0 &-1\\
                       0 &0& 1 & 0 \\
                       0 & 1& 0 & 0 \\
                      -1 & 0 & 0 &0 \\
                     \end{array}
                   \right).
\end{eqnarray*}
Hence after the spin flip operation the matrix (\ref{20}) is determined by
\begin{eqnarray*}\rho^C&=&\left(
                                 \begin{array}{cccc}
                                 0&0&0&0\\
                                  0& \rho_{33}^{\ast}& \rho_{32}^{\ast}& -\rho_{31}^{\ast}\\
                                   0&\rho_{23}^{\ast}& \rho_{22}^{\ast}& -\rho_{21}^{\ast}\\
                                   0&-\rho_{13}^{\ast}& -\rho_{12}^{\ast}& \rho_{11}^{\ast}\\
                                 \end{array}
                               \right).
                               \end{eqnarray*}
If matrix $\widetilde{\rho}$ is real, then the eigenvalues of $\widetilde{\rho}\rho^C$ are the following
        \begin{eqnarray*}\lambda_1^{C} &=&\frac{\rho_{23}^2+\rho_{32}^2+(\rho_{23}+\rho_{32})\sqrt{\rho_{23}^2-2\rho_{23}\rho_{32}+\rho_{32}^2+4\rho_{22}\rho_{33}}}{2}+\rho_{22}\rho_{33},\\
\lambda_2^{C} &=&\frac{\rho_{23}^2+\rho_{32}^2-(\rho_{23}+\rho_{32})\sqrt{\rho_{23}^2-2\rho_{23}\rho_{32}+\rho_{32}^2+4\rho_{22}\rho_{33}}}{2}+\rho_{22}\rho_{33},\\
\lambda_3^{C} &=&0,\quad
\lambda_4^{C}=0.
\end{eqnarray*}
Substituting the latter eigenvalues in (\ref{C}) we can obtain the concurrence for the qutrit state. For matrix (\ref{22}) the concurrence has the following form
 \begin{eqnarray}\label{24}C(\rho)= \max\Bigg\{0,\sqrt{p+b^2-2p^2-2b\sqrt{p-2p^2}}-\sqrt{p+b^2-2p^2+2b\sqrt{p-2p^2}}\Bigg\}.
 \end{eqnarray}
 In Figure \ref{fig:3} the concurrence (\ref{24}) is shown  for various parameters $b$.

\subsection{Symmetric spin state}
\noindent
Let us consider a composite state of two $1/2$ particles. The following basis can be obtained
\begin{eqnarray*}&&\overline{g_1}=|\uparrow\uparrow>=|1,1>, \quad \overline{g_2}=\frac{1}{\sqrt{2}}|\uparrow\downarrow>+|\downarrow\uparrow>=|1,0>,\overline{g_3}=|\downarrow\downarrow>=|1,-1>, \\
&&\overline{g_4}=\frac{1}{\sqrt{2}}|\uparrow\downarrow>-|\downarrow\uparrow>=|0,0>,
\end{eqnarray*}
where $\overline{g_1},\overline{g_2},\overline{g_3}$ correspond to symmetrical states and $\overline{g_4}$ to asymmetrical state. The connection between basis $\overline{g}$ and basis $\overline{e}$ for two-qubit system (see Section 2) is the following
\begin{eqnarray*}\overline{g_j}&=&\sum\limits_{k=1}^{4} c_{kj}\overline{e_k},\quad j=1,2,3,4,
\end{eqnarray*}
where $c_{kj}$ are  elements of matrix $C$ representing the change of the basis
\begin{eqnarray*}C&=&\left(
                                 \begin{array}{cccc}
                                   1& 0& 0&  0\\
                                   0& \frac{1}{\sqrt{2}}& \frac{1}{\sqrt{2}}&  0\\
                                   0&\frac{1}{\sqrt{2}}& -\frac{1}{\sqrt{2}}&  0\\
                                   0& 0& 0&  1\\
                                 \end{array}
                               \right).
                               \end{eqnarray*}
  The state of qutrit of the composed two spin $1/2$ system is entangled state. The two-qubit system state given by the density matrix (\ref{2})
     with specific choice of coefficients is entangled too. But these two entanglements correspond to different choices of basis in the four dimensional Hilbert space of the system states.
\par In the basis  $\overline{g}$, the $4\times4$ density matrix has the following form
\begin{eqnarray}\langle jm|\rho|j'm'\rangle={\left(
                                 \begin{array}{cccc}
                                   \rho_{11}^{11}& \rho_{10}^{11}& \rho_{1-1}^{11}&  \rho_{10}^{10}\\
                                   \rho_{01}^{11}& \rho_{00}^{11}& \rho_{0-1}^{11}&  \rho_{00}^{10}\\
                                   \rho_{31}^{11}& \rho_{32}^{11}& \rho_{33}^{11}&  \rho_{-10}^{10}\\
                                   \rho_{01}^{01}& \rho_{00}^{10}& \rho_{0-1}^{01}&  \rho_{00}^{00}\\
                                 \end{array}
                               \right)}\,, \label{8}
                               \end{eqnarray}
 where we use the notation $\rho_{mm'}^{jj'}$ for the matrix elements.
 If we cut from the matrix (\ref{8}) the fourth row and column, corresponding to asymmetrical state, and replace them  with the zero row and column, we obtain a matrix similar to
 (\ref{11}), i.e.
 \begin{eqnarray}\rho_s={\left(
                                 \begin{array}{cccc}
                                   \rho_{11}^{11}& \rho_{10}^{11}& \rho_{1-1}^{11}&  0\\
                                   \rho_{01}^{11}& \rho_{00}^{11}& \rho_{0-1}^{11}&  0\\
                                   \rho_{31}^{11}& \rho_{32}^{11}& \rho_{33}^{11}& 0\\
                                   0& 0& 0&  0\\
                                 \end{array}
                               \right)}\,. \label{9}
                               \end{eqnarray}
Hence the spin 1 system considered as symmetric
spin state of two spin $1/2$ particles can be characterized by the density matrix (\ref{9}). The inequality (\ref{10}) holds for the latter matrix.
\par The technique of appending zero rows and columns to the density matrix can be applied to the matrix (\ref{1}). From four dimensional matrix we can get $6\times6$ density matrix
\begin{eqnarray}\widetilde{\widetilde{\rho}}={\left(
                                 \begin{array}{cccccc}
                                 0&0& 0& 0& 0&0\\
                                   0&\rho_{11}& \rho_{12}& \rho_{13}& \rho_{14}&0\\
                                   0&\rho_{21}& \rho_{22}& \rho_{23}& \rho_{24}&0\\
                                   0&\rho_{31}& \rho_{32}& \rho_{33}& \rho_{34}&0\\
                                   0&\rho_{41}& \rho_{42}& \rho_{43}& \rho_{44}&0\\
                                   0&0& 0& 0& 0&0\\
                                 \end{array}
                               \right)}\,. \label{19}
                               \end{eqnarray}
As in the previous section, we can construct the reduced density matrices
                                 \begin{eqnarray*}\widetilde{\widetilde{\rho}}_1&=&\left(
                              \begin{array}{cc}
                                \rho_{11}+\rho_{22} & \rho_{14} \\
                               \rho_{41} &  \rho_{33}+\rho_{44} \\
                              \end{array}
                            \right),
                          \widetilde{\widetilde{\rho}}_2=\left(
                              \begin{array}{ccc}
                             \rho_{33} & \rho_{34}&0 \\
                              \rho_{43} & \rho_{11}+\rho_{44} & \rho_{12} \\
                               0&\rho_{21} &  \rho_{22} \\
                              \end{array}
                            \right).
\end{eqnarray*}
Therefore, the entropic inequality is
\begin{eqnarray}\label{10}&&- Tr\left(
                              \begin{array}{cc}
                                \rho_{11}+\rho_{22} & \rho_{14} \\
                               \rho_{41} &  \rho_{33}+\rho_{44} \\
                              \end{array}
                            \right)\ln\left(
                              \begin{array}{cc}
                                \rho_{11}+\rho_{22} & \rho_{14} \\
                               \rho_{41} &  \rho_{33}+\rho_{44} \\
                              \end{array}
                            \right)\\\nonumber
&-&Tr\left(
                              \begin{array}{ccc}
                             \rho_{33} & \rho_{34}&0 \\
                              \rho_{43} & \rho_{11}+\rho_{44} & \rho_{12} \\
                               0&\rho_{21} &  \rho_{22} \\
                              \end{array}
                            \right)\ln\left(
                              \begin{array}{ccc}
                             \rho_{33} & \rho_{34}&0 \\
                              \rho_{43} & \rho_{11}+\rho_{44} & \rho_{12} \\
                               0&\rho_{21} &  \rho_{22} \\
                              \end{array}
                            \right)+\sum\limits_{i=1}^{4}\widetilde{\widetilde{\lambda}}_i\ln\widetilde{\widetilde{\lambda}}_i\geq 0.
\end{eqnarray}
Thus, we can construct the matrices of any higher even dimensions from matrices of odd dimensions  and write for them the variety of entropic inequalities. The same technique
can be extended on the matrices of even dimensions as it was shown on the example (\ref{19}) to get more entropic inequalities for them.

\section{Conclusion}
\noindent
To conclude we point out the main results of our work. We considered the separability and entanglement notion for the qutrit state, e.g tree-level atom. The analog of concurrence and negativity were introduced for the non composite qutrit state. Composite system with spin $s=1$ corresponding to symmetric state of two spin $1/2$ particles was studied and the concurrence and negativity were evaluated explicitly.  Using the known property of non invariance of the entanglement with respect to change of basis of the Hilbert space of states we studied the notion of entanglement for symmetric state of spin $1/2$ particle and its difference with respect to known entanglement properties of Bell state. Applying  the known technics for the two-qubit systems the new entropic inequalities for such system were obtained.  Latter results were demonstrated on the examples of one and two parametric density matrices.

\nonumsection{Acknowledgements}
\noindent
L. A. M. acknowledges the financial support provided within the Russian Foundation for Basic Research, grant 13-08-00744 A.

\nonumsection{References}

\end{document}